\documentclass[12pt,preprint]{aastex}

\DeclareMathAlphabet{\mathbf}{OT1}{cmr}{bx}{it}
\begin{document}

\title{The Physical Origin of the Scattering
Polarization of the\break \ion{Na}{1} D-Lines
in the Presence of Weak 
Magnetic Fields}

\author{Javier~Trujillo Bueno\altaffilmark{1}}
\affil{Instituto de Astrof{\'\i}sica de Canarias, E-38200 La Laguna, 
Tenerife, Spain}
\author{Roberto Casini}
\affil{High Altitude Observatory,
National Center for Atmospheric Research,\altaffilmark{2}
\break Boulder CO 80307-3000, U.S.A.}
\author{Marco Landolfi}
\affil{Osservatorio Astrofisico di Arcetri, Largo E. Fermi 5, I-50125 Firenze,
Italy}
\author{Egidio Landi Degl'Innocenti}
\affil{Dipartimento di Astronomia e Scienza dello Spazio,
       Universit\`a di Firenze,\break
       Largo E. Fermi 2, I-50125 Firenze, Italy}
\altaffiltext{1}{Consejo Superior de Investigaciones Cient\'\i ficas, Spain}
\altaffiltext{2}{The National Center for Atmospheric Research is
sponsored by the National Science Foundation.}

\begin{abstract}
 
We demonstrate that the atomic alignment of the hyperfine-structure components
of the ground level S$_{1/2}$
of \ion{Na}{1} and of the upper level P$_{1/2}$ of the D$_1$ line
are practically negligible for magnetic strengths $B>10\,\rm G$, and
virtually zero for $B\ga 100\,\rm G$. This occurs independently 
of the magnetic-field inclination on the stellar surface (in 
particular, also for vertical fields). Consequently, the 
characteristic antisymmetric linear-polarization signature
of the scattered light 
in the D$_1$ line is practically suppressed in the presence of 
magnetic fields larger than 10~G, regardless of their inclination.
Remarkably, we find that the scattering
polarization amplitude of the D$_2$ line increases steadily with the
magnetic strength, for vertical fields above 10~G, while the 
contribution of alignment to the polarization of the D$_1$ line
rapidly decreases. Therefore, we suggest that spectropolarimetric 
observations of the ``quiet'' solar chromosphere showing significant 
linear polarization peaks in both D$_1$ and D$_2$ cannot be 
interpreted in terms of one-component magnetic field models, implying that the 
magnetic structuring of the solar chromosphere could be substantially 
more complex than previously thought. 

\end{abstract}

\subjectheadings{atomic processes --- polarization --- scattering --- 
stars: magnetic fields --- Sun: chromosphere}

\section{Introduction}
\label{sec:intro}

In a recent work, one of the authors (\citealt{LA98})
concluded that his 
explanation in terms of ground-level atomic polarization of the 
``enigmatic'' linear polarization peaks of the \ion{Na}{1} D-lines, 
observed by \citet{STEN97} in ``quiet'' regions close to 
the solar limb, implies that the magnetic field in the lower solar 
chromosphere must be either isotropically distributed and extremely 
weak (with $B\la 0.01\,\rm G$) or, alternatively, practically radially 
oriented. 
That investigation was based on a formulation of 
line scattering polarization that is valid in the absence of magnetic fields.
The suggestion that the magnetic field of the lower solar chromosphere
cannot be stronger than about 0.01 G unless it is oriented
preferentially along the radial direction was based on the sizeable 
amount of ground-level polarization required to fit the $Q/I$ 
observations of \citet{STEN97}, and on the assumption that
the atomic polarization of the ground-level of \ion{Na}{1} must be
sensitive to much weaker magnetic fields than the atomic polarization
of the upper levels of the D$_1$ and D$_2$ lines.

On the whole, Landi Degl'Innocenti's (\citeyear{LA98}) argument that the
observed linear polarization peaks in the cores of the \ion{Na}{1}
D-lines are due to the presence of ground-level atomic polarization
seems very convincing. However, for a rigorous interpretation 
of spectropolarimetric observations
(e.g., \citealt{VMP01,STEN01}) it is of fundamental importance
to clarify the physical origin of this polarization by carefully 
investigating how it is actually produced, and modified by the 
action of a magnetic field of given strength and 
inclination.\footnote{Remarkably, some useful information
can be found in the atomic physics literature, notably in the paper by 
\citet{ELL34} regarding their determination of
hyperfine separation constants, and in the work of \citet{LEH69} 
concerning the orientation of the diamagnetic ground state of Cadmium
by optical pumping.} 

\section{Formulation of the problem}
\label{sec:formulation}

In this Letter we shall focus  
on the ``solar prominence case'', in which a slab of 
solar chromospheric plasma at 6000 K, situated at 10 arcsec 
($\approx 7000\,\rm km$) above the visible solar limb, 
and permeated by a
magnetic field of given strength and orientation,
is illuminated from below (hence, anisotropically) by
the photospheric radiation field, which is
assumed to be unpolarized 
and with rotational symmetry around the
solar radial direction through the scattering point.
The degree of anisotropy of the incident radiation field
is calculated as in Landolfi \& Landi Degl'Innocenti (1985, hereafter
LL85), using the limb-darkening data for the Na {\sc i} D-lines
from Pierce \& Slaughter (1982), and a Gaussian absorption profile with
${\Delta\lambda}_{\rm D}=41\,\rm{m}\rm\AA$ (corresponding to $T=6000\,\rm K$).
The resulting anisotropy factors for the two lines are
$w({\rm D}_1)=0.126$ and $w({\rm D}_2)=0.118$, where 
$w=\sqrt{2}\,\bar{J}^2_0/\bar{J}^0_0$
(with $\bar{J}^K_Q$ the radiation field tensors;
see, for example, Trujillo Bueno 2001
and note that $-1/2\,{\la}\,w\,{\la}\,1$). 

In order to investigate this problem, we have applied
the quantum theory of spectral-line polarization
in the limit of complete frequency redistribution,
and in the collisionless regime,
as developed by \citet{LA83}.
The excitation of the atomic system is described by a 
set of $\rho^K_Q$ elements, which are the irreducible 
spherical tensors 
of the atomic density matrix (e.g., the review by
\citealt{JTBA01}). 
We adopt a three-level model 
of \ion{Na}{1} consisting of
the ground level ($3\,{}^2{\rm S}_{1/2}$), the upper level of 
the D$_1$ line ($3\,{}^2{\rm P}_{1/2}$), and the upper level
of the D$_2$ line ($3\,{}^2{\rm P}_{3/2}$). We also take into 
account the hyperfine structure (HFS) of Sodium,  
due to its nuclear spin with $I=3/2$.

We describe the Zeeman splittings of the \ion{Na}{1} levels
most generally in terms of the incomplete Paschen-Back effect.
In fact, as shown, e.g., in Fig.~5 of LL85,
in the 10--50 G range numerous level crossings occur among the magnetic 
sublevels of the HFS levels
with $F=1,2,3$ of P$_{3/2}$, whereas fields with strengths in
the kilogauss range are necessary to produce level crossing
within the HFS levels with $F=1,2$ of S$_{1/2}$. Therefore,
besides population imbalances and quantum interferences 
(or coherences) between the magnetic sublevels of each $F$-level,
we must also take into account coherences between magnetic
sublevels pertaining to different $F$-levels within a given $J$-level.
Instead, we neglect coherences between the two $J$-levels
P$_{1/2}$ and P$_{3/2}$, since they are presumably of secondary 
importance for the generation of the line-core polarization peaks,
given the sizeable energy separation between the upper levels 
of the D$_1$ and D$_2$ lines.
Our model atom then implies 384 $^{JI}\rho^K_Q(F,F')$ elements
(with $K=|F-F'|,\ldots,F+F'$ and $Q=-K,\ldots,K$), which are the
unknowns of the linear system representing the statistical
equilibrium problem for \ion{Na}{1}. In this Letter, these
quantities are calculated in a 
reference frame with the $z$-axis (i.e., the quantization axis)
along the solar radial direction through the scattering point.

In summary, our approach is similar to that of LL85,
but with the following fundamental improvement. In the 
expressions of the Stokes 
components of the emission vector 
($\epsilon_I,\epsilon_Q,\epsilon_U,\epsilon_V$)
we now take fully into account the energy separation of the 
various HFS components of the D$_1$ and D$_2$ lines, along 
with their Zeeman splittings in the presence of the external 
magnetic field. This is crucial in order
to obtain non-zero linear polarization for the \ion{Na}{1} D$_1$ line.

\section{Polarizability of the \ion{Na}{1} levels}
\label{sec:polar}

We have solved numerically the linear system of 384 equations
in the unknowns $\rho^K_Q(F,F')$ mentioned before, for magnetic 
strengths between 0 and 1000~G, and for various inclinations
($\vartheta_B$) of the magnetic field vector from the
solar vertical. The eight $\rho^0_0(F,F)$ elements quantify the 
populations of the various $F$-levels, and they produce the 
dominant contribution to the emergent Stokes-$I$ parameter.
The $\rho^2_Q$ elements (the {\em alignment} components)
contribute to the {\em linear} polarization signals, which we 
quantify by the Stokes parameters $Q$ and $U$. The $\rho^1_Q$ 
elements (the {\em orientation} components) contribute to the 
{\em circular} polarization of the scattered 
radiation. (We recall that in an aligned atomic system, states 
of different $|M_F|$ are unequally populated, while the populations
in $M_F$ and $-M_F$ are the same. In contrast, an oriented
system is characterized by different populations in the $M_F$ and
$-M_F$ states. We are dealing here with
an atomic system which is both aligned
and oriented).
The contributions from the longitudinal and transverse Zeeman 
effects are also accounted for, although they become dominant
only for relatively strong fields.

\begin{figure}[t]
\epsscale{1}
\plotone{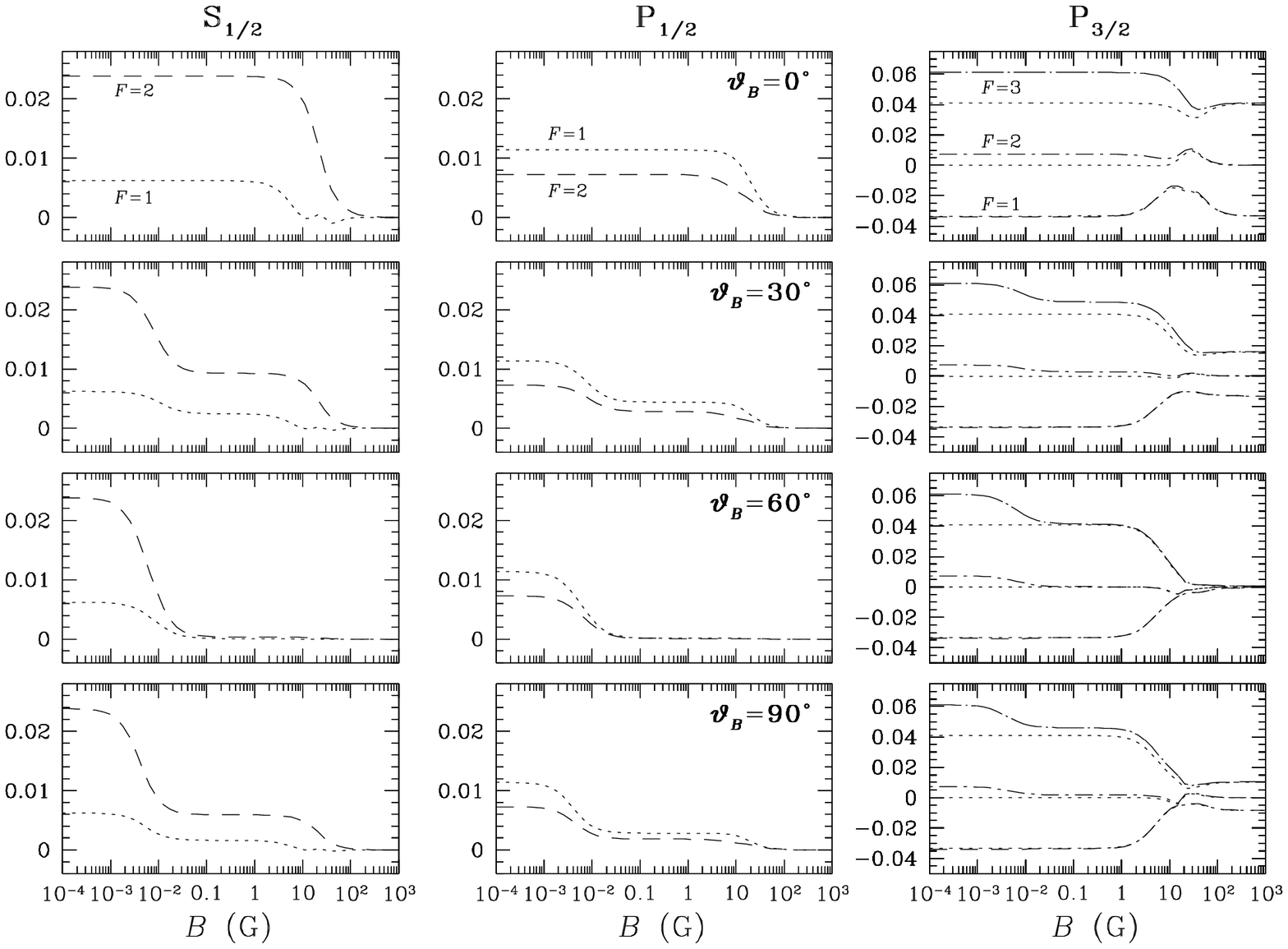}
\caption{\label{fig:alignment}
The fractional atomic alignment, 
$\sigma^2_0$,  
in the first three levels of \ion{Na}{1} as a function of the 
magnetic strength and for various inclinations ($\vartheta_B$) 
of the magnetic field vector. 
All quantities are referred to a
reference frame with the $z$-axis (i.e., the quantization axis)
along the solar vertical through the scattering point.
The dotted lines 
in the panels corresponding to the level P$_{3/2}$ 
show the $\sigma^2_0$ values assuming a completely 
unpolarized ground level.
Note that, even for vertical fields, atomic depolarization of the lower 
and upper levels of the D$_1$ line is total when the level P$_{3/2}$ is in the 
regime of complete Paschen-Back effect.}
\end{figure}

Given that we are interested in understanding the generation of 
{\em linear} polarization signals in the presence of weak
magnetic fields, we focus here on the {\em alignment} components. 
In Fig.~\ref{fig:alignment}, for each $F$-level, we show
$\sigma^2_0(F)={\rho^2_0(F,F)/\rho^0_0(F,F)}$,
which quantifies the fractional {\em population imbalance} of
the level. Since the spectral dependence of the incident
radiation field is practically negligible over the frequency intervals
encompassing the Zeeman components of each of the two spectral lines
({\em flat-spectrum approximation\/}), a necessary condition for
inducing atomic alignment by means of an unpolarized
radiation field is that the illumination of the atomic system be anisotropic.
Moreover, atomic orientation can only be originated through
the alignment-to-orientation conversion mechanism
discussed by \citet{KEMP84}.

Figure~\ref{fig:alignment} shows the sensitivity of $\sigma^2_0(F)$
to the magnetic field strength and inclination.
First of all, we note
that the largest values are obtained for the level
P$_{3/2}$, which can carry atomic alignment even neglecting HFS. On the
contrary, both the lower and upper levels of the D$_1$
line, with electronic angular momentum $J=1/2$, can carry atomic
alignment only because of HFS, as each of these levels splits into
two {\em polarizable} HFS levels with $F=1$ and $F=2$. However,
it is found that only the level ${\rm P}_{3/2}$ can be polarized
directly via the anisotropic illumination. The levels of the D$_1$ line,
instead, are directly sensitive only to radiation intensity, but
they nonetheless become polarized when the atomic polarization
of the level P$_{3/2}$ is transferred to the level S$_{1/2}$
via spontaneous emission in the D$_2$ line, and then from the level
S$_{1/2}$ to the level P$_{1/2}$ via radiative absorption in the
D$_1$ line, in a process known as {\em repopulation pumping}
(e.g., \citealt{JTBA01}).
In fact, this explains one of the various remarkable features
of Fig.~\ref{fig:alignment}, i.e., the fact that the atomic alignment in the lower and
upper levels of the D$_1$ line are equally sensitive to the magnetic
strength, independently of the magnetic field inclination. 

For instance, we see that a {\em non-vertical} magnetic field 
of the order of 0.01 G is sufficient to produce a serious reduction of
the atomic alignment of both the lower and upper levels of the D$_1$ line.
This is due to Hanle depolarization of the S$_{1/2}$ ground level, which
occurs when the Larmor frequency corresponding to the magnetic
field becomes comparable to the inverse lifetime for
radiative absorption of that level.  
Nonetheless, the alignment of the
level $F=2$ of S$_{1/2}$ 
is still significant for fields
up to 10~G, except for field inclinations close to the Van
Vleck angle ($\vartheta_B=54.73^\circ$). 

For non-vertical fields the alignment of the level P$_{3/2}$
is also sensitive to magnetic strengths between 0 and 10~G,
but the depolarization takes
place rather smoothly. An interesting point to note here is
the sizeable feedback of the
ground-level polarization on the alignment of the $F$-levels of
P$_{3/2}$, with the exception of the level $F=1$.
(This behavior can be understood analytically via
inspection of the corresponding transfer rates).
As previously indicated, such a feedback takes place because
the upper level of the D$_2$ line can be repopulated
as a result of absorptions from the {\em polarized} ground level. 

The most remarkable feature of
Fig.~\ref{fig:alignment} is that, independently of
the magnetic field inclination
(e.g., even for a purely vertical magnetic field),
the atomic alignment of each of the two levels involved in the
D$_1$ line transition is suddenly
reduced for magnetic strengths larger than
10~G, and practically vanishes for strengths larger than 100~G.
We stress the fact that this depolarization is {\em not} due
to the Hanle effect, as it occurs also for vertical fields.
A thorough investigation of this phenomenon shows that the
vanishing of atomic alignment in the
levels with $J=1/2$ sets in when the electronic and nuclear
angular momenta, $\mathbf J$ and $\mathbf I$, are decoupled,
for the atom in the excited state P$_{3/2}$. In the case of
\ion{Na}{1}, this decoupling is reached in the limit
of the complete Paschen-Back effect of the level P$_{3/2}$,
i.e., for magnetic strengths $B\ga 100\,\rm G$. In such regime,
it is found that the transfer of atomic alignment from the
level ${\rm P}_{3/2}$ to the ground level is inhibited. At
the same time, the alignment of the level $F=2$ of P$_{3/2}$
must vanish as well. The analytical proof of these properties
will be given in a forthcoming paper (Casini et al.~2001,
in preparation).

It is also of interest to note that the
repopulation pumping process works efficiently in \ion{Na}{1} thanks to the
fact that the HFS of the level ${\rm P}_{3/2}$
is of the same order of magnitude of its natural width.
If the frequency intervals between the HFS levels of P$_{3/2}$
were instead substantially smaller than the natural width of this level,
then we would have a negligible HFS interaction during the
lifetime of the level ${\rm P}_{3/2}$, with the result of a drastic
reduction in the efficiency of the repopulation pumping process that polarizes
the ground level of Sodium, regardless of the magnetic field strength.

\section{Observable effects of the atomic alignment}
\label{sec:emergent}

The theory of the Hanle effect for a two-level atom
devoid of HFS
predicts no modication of the emergent linear polarization
with increasing strength
of a magnetic field
oriented parallel to the symmetry axis of the
incident radiation (e.g., \citealt{LA85}).
For this reason, and given the
conclusions of the previous section, it is of great interest
to investigate
the emergent polarization of the \ion{Na}{1} D-lines
for 90$^\circ$ scattering events
as a function of the strength of a vertical magnetic field. For this
case, and choosing the reference direction for positive Stokes $Q$ 
parallel to the limb, the only non-zero Stokes parameter is $Q$.

Figure~2 shows profiles of $Q/I_{\rm max}$, where $I_{\rm max}$
indicates the peak intensity of the emission line,
for magnetic strengths between 0 and 100~G.
The top panels refer to the solution of the statistical
equilibrium problem outlined in \S~\ref{sec:polar}.  
The results shown in the two bottom panels,
instead, are obtained assuming that the ground level of 
\ion{Na}{1} is totally unpolarized (e.g., by the presence of 
depolarizing collisions).

The first interesting feature of the
D$_2$ line polarization
is the {\em increase} of the linear
polarization degree as the strength of the
vertical magnetic field increases beyond 10 G.
This is caused by the interferences, $\rho^2_Q(F,F')$ (not shown
in Fig.~\ref{fig:alignment}), of the HFS levels in the P$_{3/2}$ level. 
Our calculations for {\em inclined} fields with $B\la 100\,\rm G$ 
(not given here)
show first a decrease and then an increase of the linear polarization,
but the maximum polarization amplitude of the D$_2$ line corresponds to the 
$B=0\,\rm G$ case. In general, the
polarization of both D$_1$ and D$_2$ fluctuates
significantly with strength and inclination.

\begin{figure}[t]
\epsscale{1}
\plotone{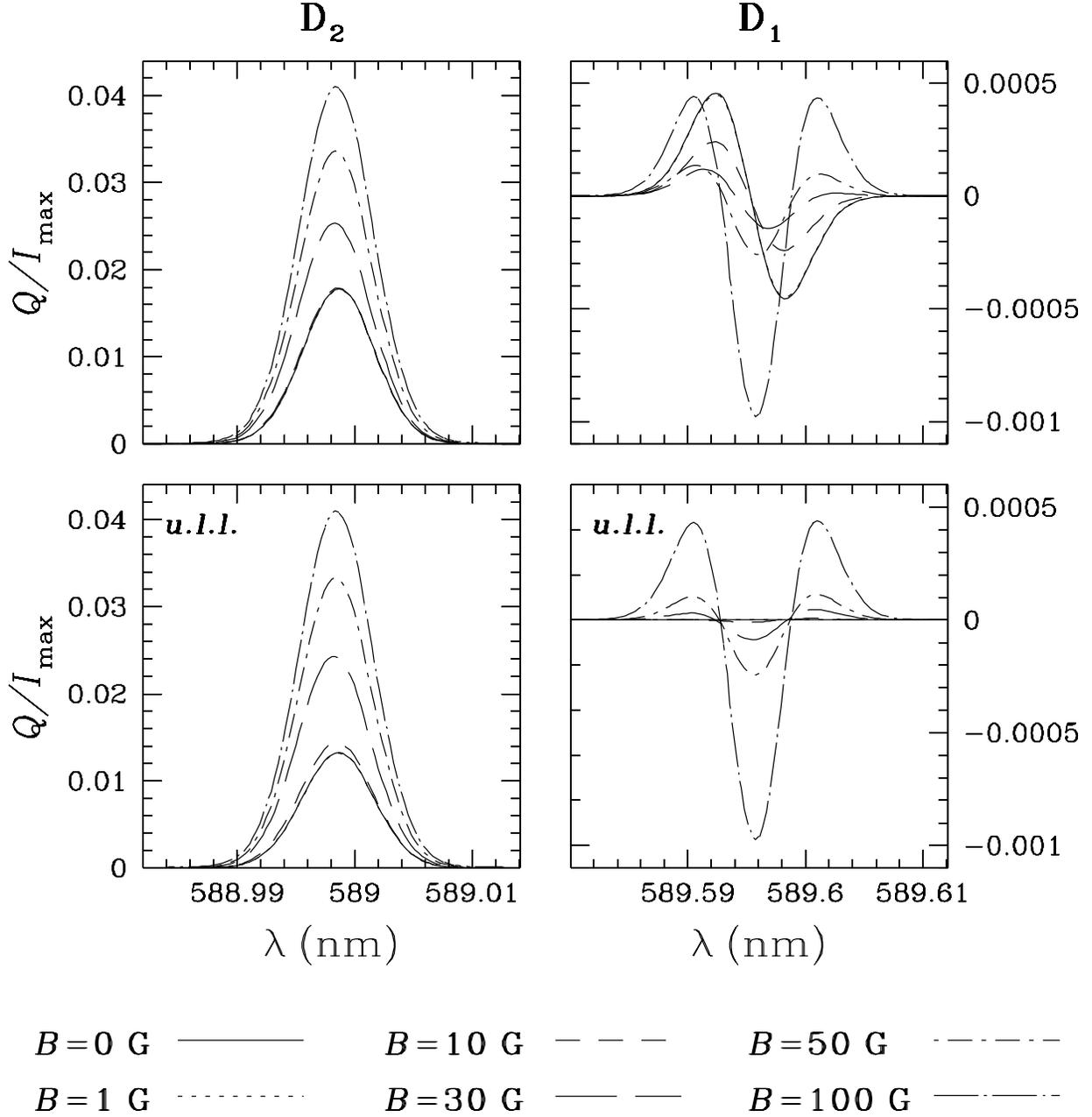}
\figcaption{\label{fig:profiles}
The emergent linear polarization of the \ion{Na}{1} D-lines
in a $90^\circ$ scattering event
for increasing values of the magnetic strength of the
assumed vertical field. The positive reference direction for 
Stokes $Q$ is along the line perpendicular to the radial 
direction through the scattering point. The top panels take 
into account the feedback of ground level polarization on 
the atomic polarization of the two upper levels. The
two panels with the label ``$u.l.l$'', instead, neglect the 
influence of ground-level polarization. 
The kinetic temperature is 6000 K. We point out that 
the amplitude of the D$_1$ peaks is 
particularly sensitive to the Doppler width.}
\end{figure}

We note that the polarization of the D$_2$ line
is practically unaffected by ground-level polarization
for fields larger than 10~G. As shown in Fig.~\ref{fig:alignment},
for weaker fields the existing ground-level polarization
has a significant feedback on the alignment of the ${\rm P}_{3/2}$ level,
which in turn produces a significant but small enhancement
of the emergent linear polarization in the D$_2$-line core,
with respect to the case of totally unpolarized ground level (see Fig.~\ref{fig:profiles}).

As shown in Fig.~\ref{fig:profiles}, for magnetic strengths $B\la 10\,\rm G$,
the emergent linear polarization of the
D$_1$ line owes its very existence
to the presence of ground-level polarization.
Note that its Stokes $Q$ profile is 
{\em antisymmetric}.\footnote{This peculiar shape has
been observed by \citet{JTBB01} in ``quiet''
regions close to the solar limb, as shown in Fig.~\ref{fig:alignment} of
\citet{JTB01}. See also \citet{BOM02}.}
In Fig.~\ref{fig:profiles} we can see that
a 1~G {\em vertical} field has practically no 
effect on the emergent linear polarization,
while a 10~G {\em vertical} field reduces the signal by a factor of two,
as a result of the significant reduction of the atomic alignment 
of the HFS levels of the upper level P$_{1/2}$ (see Fig.~\ref{fig:alignment}). 
What we see for stronger fields in the corresponding
top panel of Fig.~\ref{fig:profiles} is the result
of the combined action of the atomic
alignment of the upper level of the D$_1$ line and of
the transverse Zeeman effect.
Since the Zeeman splittings for the magnetic strengths of interest
are determined
by the incomplete Paschen-Back effect,
and the \ion{Na}{1} D-lines are the combination of transitions
between $F$-levels weighted by different Land\'e $g$-factors,
the Stokes $Q$ profiles can be asymmetric, even if the
ground level is forced to be totally unpolarized (see bottom panels of Fig.~\ref{fig:profiles}).
Nonetheless, it can be analitically proven that
the wavelength integrated Stokes $Q$ parameter of D$_1$
is always zero (see LL85).
It is also of interest to point out that, if the
field is horizontal with randomly distributed azimuth,
then the Stokes $Q$ amplitude of the D$_1$ line is reduced
by a factor of four for a magnetic strength of 1~G, with respect to the 
non-magnetic case. 
 
In general, the linear polarization of the D$_1$ line is dominated
by the transverse Zeeman effect for magnetic strengths 
$B\ga 50\,\rm G$ (the alignment of the D$_1$ levels is 
practically zero for such field intensities).
For example, from Fig.~\ref{fig:profiles} we see that the amplitude of the 
Stokes $Q$ signal 
in the presence of a {\em vertical} field with $B\approx 100\,\rm G$
is comparable to the amplitude produced by scattering processes in the 
absence of magnetic fields, but the linear polarization signature has a
profile characteristic of the transverse Zeeman effect, with a 
polarization peak at the line core. 
On the other hand, the Stokes $Q$ signal in the D$_2$ line 
for $B\approx 100\,\rm G$ is still ``scattering like,''
being determined essentially by the contribution of the 
alignment of the level ${\rm P}_{3/2}$. In fact, for this line, 
the signature of the transverse Zeeman effect begins to appear 
only for fields $B\ga 500\,\rm G$.

\section{Concluding remarks}
\label{sec:conclu}

One of the most interesting results of this investigation is that
the atomic polarization of the HFS levels of the ${\rm S}_{1/2}$ 
and ${\rm P}_{1/2}$ states of \ion{Na}{1} is practically negligible 
for $B > 10\,\rm G$, and virtually vanishes for 
$B {\ga} 100\,\rm G$, even for a purely vertical field.
Consequently, the characteristic antisymmetric 
scattering polarization signature of the D$_1$ line 
is practically suppressed in 
the presence of fields larger than 10~G, regardless of 
their inclination.

Concerning the observable effects, we find that the scattering
polarization amplitude of the D$_2$ line increases steadily with
the magnetic strength, in the case of vertical fields larger
than 10~G, whereas the contribution of atomic alignment to 
the linear polarization of the D$_1$ line rapidly decreases. 
On the contrary, for vertical fields such that $B\la 10\,\rm G$ 
(or, alternatively, for turbulent or canopy-like fields with a 
predominance of much weaker fields) it is possible to have a 
non-negligible scattering polarization signal for the D$_1$ line, 
but then the maximum D$_2$ core amplitude corresponds
to the $B=0\,{\rm G}$ case.
From this we tentatively conclude that spectropolarimetric 
observations of the ``quiet'' solar chromosphere 
showing significant scattering polarization peaks in both the D$_1$ 
and D$_2$ line cores cannot be 
interpreted in terms of one-component magnetic field models, 
suggesting that the magnetic structuring of the solar 
chromosphere could be substantially more complex than 
previously thought. For instance, in the presence of a topologically complex
distribution of ``weak'' solar magnetic fields, 
the D$_2$ line core would respond mainly to the 
strongest and preferentially radially oriented
fields, while the D$_1$ line to the weakest and more randomly
oriented fields. It remains to be seen whether or not 
this conclusion is validated after we take 
fully into account radiative transfer effects and
the role of {\em dichroism} (\citealt{JTB97})
on the emergent polarization
of the ``enigmatic'' Na {\sc i} D-lines. 

\acknowledgements
One of the authors (J.T.B.) is grateful to the
High Altitude Observatory (U.S.A.) and to 
the University of Florence (Italy)
for short-term visitor grants, which have
facilitated this collaboration. This work
has been partly funded by the Spanish 
PNAYA through project AYA2001-1649.

\newpage

\newpage

\end{document}